\RequirePackage{ifpdf}
\ifpdf
	\documentclass[twocolumn, showkeys, pdftex]{bioeng}
\else
	\documentclass[twocolumn, showkeys, dvips]{bioeng}
\fi

\usepackage{amsmath, amsthm}
\RequirePackage{ifpdf}
\ifpdf
  \usepackage[pdftex]{graphicx}
  \graphicspath{{pdf-figures/}}
  \usepackage[%
    pdftex,%
    pdftitle = {Fermentation\ kinetics\ including\ product\ and\ substrate\ inhibitions\ plus\ biomass\ death:\ a\ mathematical\ analysis},%
	pdfauthor = {Mathieu\ Bouville},%
	pdfsubject = {Writing\ fermentation\ kinetics\ equations\ in\ phase\ space},%
	pdfkeywords = {fermentation\ kinetics;\ model;\, inhibition},%
	pdfpagemode = UseOutlines,%
    hyperindex,%
    bookmarksopen,%
    bookmarksopenlevel=1,%
  ]{hyperref}
\else
  \usepackage[dvips]{graphicx}
  \usepackage[%
    dvips,%
    pdftitle = {Fermentation\ kinetics\ including\ product\ and\ substrate\ inhibitions\ plus\ biomass\ death:\ a\ mathematical\ analysis},%
	pdfauthor = {Mathieu\ Bouville},%
	pdfsubject = {Writing\ fermentation\ kinetics\ equations\ in\ phase\ space},%
	pdfkeywords = {fermentation\ kinetics;\ model;\, inhibition},%
	pdfpagemode = UseOutlines,%
    hyperindex,%
    bookmarksopen,%
    bookmarksopenlevel=1,%
  ]{hyperref}
\fi

\newcommand{\rmd}{\mathrm{d}}
\newtheorem{thm}{Theorem}

\begin{document}

\title[M. Bouville \:---\: Fermentation kinetics with inhibitions, biomass death]{Fermentation kinetics including product and substrate inhibitions\\ plus biomass death: a mathematical analysis}

\author{Mathieu Bouville}
	\email{m-bouville@imre.a-star.edu.sg}
	\affiliation{Institute of Materials Research and Engineering, Singapore 117602}

\date{\today}

\begin{abstract}
Fermentation is generally modelled by kinetic equations giving the time evolutions for biomass, substrate, and product concentrations. Although these equations can be solved analytically in simple cases if substrate/\linebreak[2]product inhibition and biomass death are included, they are typically solved numerically. We propose an analytical treatment of the kinetic equations ---including cell death and an arbitrary number of inhibitions--- in which constant yield needs not be assumed. Equations are solved in phase space, i.e.\ the biomass concentration is written explicitly as a function of the substrate concentration.
\end{abstract}


\maketitle

\section*{Introduction}
Several models have been proposed to describe the kinetics of fermentation, giving the time evolutions for microbial mass $X$, substrate $S$, and product $P$. \citet{Marin-99} and \citet{Mitchell-04} have recently reviewed them.
Kinetic equations are generally of the form
\begin{align}
	\frac{\rmd S}{\rmd t}&=-\frac{1}{Y}\frac{\rmd X}{\rmd t}-m\,X, \label{dS_dt}\\
	\frac{\rmd X}{\rmd t}&=\mu\frac{S}{S+K_1}X.\label{dX_dt}
\end{align}
\noindent where $m$ is the maintenance coefficient \cite{Acevedo-87,Hill-90}, $\mu$ is the \emph{maximum} specific growth rate and $K_1$ is Monod's constant \cite{Monod-41}. Equation~(\ref{dX_dt}), originally proposed by \citet{Monod-41}, can be modified to account for product inhibition \cite{Bazua-77,Holzberg-67, Aiba-68, Ghose-79-II,Hoppe-82,Hill-90}, substrate inhibition \cite{
Ough-63, Casey-86, Ghose-79-II}, and biomass death \cite{Bailey-77, Caro-91}. Other models are possible; \citet{Cramer-01} for instance did not correlate ethanol production to biomass growth.

The biomass yield, $-\rmd X/\rmd S$, is often assumed to be constant \cite{
Ghose-79-II,Bazua-77,Cysewski-76,Cysewski-77,Hoppe-82,
Maiorella-83}. In this case $X-X^0=Y\left(S-S^0\right)$ where $X^0$ and $S^0$ are the initial values of $X$ and $S$. Equation~(\ref{dX_dt}) then gives
\begin{equation*}
	\mu\,t=\ln \frac{X}{X^0} + \frac{K_1}{S^0+X^0/Y} \left(\ln \frac{X}{X^0} + \ln \frac{S^0}{S}\right).
\label{t}
\end{equation*}

However biomass yield is not necessarily constant. \citet{Thatipamala-92} for instance found that it decreased from 0.16 to 0.03 when ethanol increased from 0 to 107 g/L. In such cases the time dependences of $S$ and $X$ are not obtained analytically, instead Eqs.~(\ref{dS_dt}) and (\ref{dX_dt}) are typically solved numerically. However since monitoring of the biomass concentration can be used to follow the evolution of fermentations an explicit expression linking $X$ and $S$ could be very useful.

\section*{Complete kinetic equation}

Equation~(\ref{dX_dt}) assumes that the microorganisms are not affected by their environment. In actuality fermentation can be slowed by large amounts of substrate (substrate inhibition) or of product (product inhibition). Biomass death may also occur. In this section we write a more complete version of Eq.~(\ref{dX_dt}) in order to take these effects into account.

Both substrate and product inhibitions can be modelled by an equation of the form 
\begin{equation}
	\frac{\rmd X}{\rmd t}=\frac{\mu\,\mu'}{K_1}\frac{S}{\displaystyle\prod_{i=1}^{n_\text{I}}\left(1+\dfrac{S}{K_i}\right)}X.
	\label{dX_dt-general}
\end{equation}
\noindent The \{$K_i$\} can be either positive or negative depending on the nature of the inhibition ($K_1$ is always positive, it is Monod's constant). They can refer to different kinds of sugars, nutrients (e.g., nitrogen) or products. $\mu'/\prod_{i=1}^{n_\text{I}}(1+S/K_i)$ must always be smaller than 1 since inhibition slows growth and it is equal to 1 in the absence of inhibition. 
$\mu'$ is given by
\begin{equation*}
\mu'=\prod_{k_i<0} \left(1+\frac{1+p_i^0}{k_i}\right)\!,
\end{equation*}
\noindent where $p_i^0=P_i^0/(\alpha_i S^0)$ and $k_i=K_i/S^0$. The product is over all $i$ corresponding to product inhibition. $\alpha_i$ is the stoichiometric coefficient for product $P_i$. 
${(1+S/K_i)}(1+S/K_j)$ can refer to two different substrates or products but it can also refer to a single variable if it is found that $(1+S/K_i)(1+S/K_j)$ can provide a better fit to the data than $1+S/K_i$ alone.

Inhibitions can slow fermentation but they cannot make the biomass decrease. In order to account for a decrease in viable biomass, cell death must be included in the model. 
Equation~(\ref{dX_dt}) becomes \cite{Bailey-77}
\begin{equation}
	\frac{\rmd X}{\rmd t}=\mu\frac{S}{S+K}X-\frac{P}{K_\text{D}}X
\label{dX_dt-death}
\end{equation}
\noindent where $K_\text{D}$ is a constant (g/L~h). The first term is biomass growth and the second term is biomass death. The way Eq.~(\ref{dS_dt}) was written $\rmd X/\rmd t$ was the same as biomass growth. But in order to use Eq.~(\ref{dX_dt-death}) one needs to distinguish between the change in the biomass due to growth (which will result in sugar consumption) and the effect of death (which has no effect on $S$). $\rmd S/\rmd t$ depends only upon the growth part of $\rmd X/\rmd t$.

The complete equation accounting for inhibitions and cell death is obtained by merging Eqs.~(\ref{dX_dt-general}) and~(\ref{dX_dt-death})
\begin{equation}
	\frac{\rmd X}{\rmd t}=\frac{\mu\,\mu'}{K_1}\frac{S}{\displaystyle\prod_{i=1}^{n_\text{I}}(1+S/K_i)}X
-\rho\,\mu\sum_{i=1}^{n_\text{D}} \frac{p_i}{\kappa_i}X,
\label{dX_dt-death_inhib}
\end{equation}
\noindent where $\kappa_i=m\,Y\,K_\text{D}(i)/(\alpha_i\,S^0)$.

\section*{Solution in phase space}
\begin{figure}
\centering
	\includegraphics[width=7.5cm]{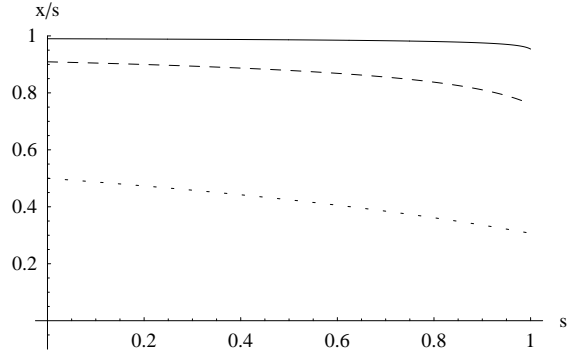}
\caption{\label{fig-dX} ${x}/{s}$ as a function of ${s}$ for $\psi=10^{-2}$ (solid line), $\psi=10^{-1}$ (dashed line), and $\psi=1$ (dotted line).}
\end{figure}

In this section we will show how one can obtain $X$ as a function of $S$ from Eqs.~(\ref{dS_dt}) and (\ref{dX_dt-death_inhib}), without assuming constant yield. But first we will proceed from Eqs.~(\ref{dS_dt}) and (\ref{dX_dt}) in order to describe the procedure on a simpler example:
\begin{equation*}
\frac{\rmd S}{\rmd X} = \frac{\rmd S/\rmd t}{\rmd X/\rmd t} =
	-\frac{1}{Y} - \frac{m}{Y}\frac{S+K_1}{S}
\end{equation*}
\noindent which gives
\begin{equation}
-\frac{1+\rho}{Y}\frac{\rmd X}{\rmd S} = 1-\frac{S^0 \psi}{S+S^0 \psi}.
\label{dX_dS-simple}
\end{equation}
\noindent The dimensionless constants $\rho$, $k_1$, and $\psi$ are
\begin{equation}
	\rho	= \frac{m\,Y}{\mu}, \quad 
	k_1		= \frac{K_1}{S^0}, \quad \text{and} \quad 
	\psi	= \frac{\rho}{1+\rho}k_1.
\label{def-sigma}
\end{equation}
\noindent Letting
\begin{equation}
	{x} = \frac{X-X^0}{Y\,S^0}\left(1+\rho\right) \quad
	\text{and} \quad
	{s} = \frac{S^0-S}{S^0},
\label{def-delta}
\end{equation}
\noindent we obtain
\begin{equation}
	{x} = {s} +\psi \ln \left(1-\frac{{s}}{1 + \psi}\right)\!.
\label{dS_dX-2}
\end{equation}
\noindent ${x}$ and ${s}$ are normalized biomass growth and substrate consumption respectively. They are zero at $t=0$ and they go to 1 when $t \to \infty$. ${s}$ depends only on ${x}$ and on $\psi$.
Figure~\ref{fig-dX} shows ${x}/{s}$ as a function of ${s}$ for three values of $\psi$. When $\psi \ll 1$ (solid line) ${x} \approx {s}$, but the assumption of constant yield is less good for larger values of $\psi$.

A similar result can be obtained from Eqs.~(\ref{dS_dt}) and (\ref{dX_dt-death_inhib}), accounting for product and substrate inhibitions combined with cell death. Equation~(\ref{dX_dS-simple}) becomes
\begin{equation}
	\frac{1}{Y}\frac{\rmd X}{\rmd S}=
		\sum_{i=1}^{n_\text{D}} \frac{p_i}{\kappa_i}
		-\frac{S+S\displaystyle\sum_{i=1}^{n_\text{D}} \frac{p_i}{\kappa_i}}{S+\displaystyle\frac{\rho\,K_1}{\mu'}\prod_{i=1}^{n_\text{I}}\left(1+\frac{S}{K_i}\right)},
\label{dX_dS-death_inhib}
\end{equation}
\noindent where $p_i = P_i/(\alpha_i S^0)=p_i^0+1-S/S^0$. Whereas Eq.~(\ref{dX_dS-simple}) was easily integrated, Eq.~(\ref{dX_dS-death_inhib}) is troublesome. The fraction must be transformed into a sum of terms of the form $1/(S-S^*)$.

\begin{thm}
Let $n\in I\!\!N,\,n \ge 2$. Let $\{\sigma_i\}\in I\!\!R^n$. Let $Q$ a polynom of degree strictly smaller than $n$.
\begin{equation}
\frac{Q(s)}{\displaystyle\prod_{j=1}^n(s-\sigma_j)} = 
\sum_{i=1}^n \frac{Q(\sigma_i)}{(s-\sigma_i)\displaystyle\prod_{\begin{subarray}{c}j=1\\j\ne i\end{subarray}}^n (\sigma_i-\sigma_j)}.
\label{theorem1}
\end{equation}
\end{thm}

\begin{proof}
Let $\Omega={\{j\in I\!\!N^*,\,j\le n\}}$. Let $\{R_i\}_{i\in\Omega}$ the polynoms defined over $I\!\!R$ by $R_i(s)=\prod_{j\ne i}{(s-\sigma_j)}$.\rule[-1.5ex]{0ex}{4ex}
Let ${(i,\,\ell)\in \Omega^2}$. $R_i(\sigma_\ell)/R_i(\sigma_i) = 1$ if $i=\ell$ and 0 otherwise. Let $T$ the polynom defined over $I\!\!R$ by $T(s) = Q(s)-\sum_{i=1}^n {Q(\sigma_i)R_i(s)}/R_i(\sigma_i)$\rule[-1.5ex]{0ex}{4ex}. ${\forall \:\ell\in\Omega}$, ${T(\sigma_\ell)=0}$. $T$ is of degree smaller than or equal to $n-1$ and it has $n$ roots. It is the zero polynom. Hence Eq.~(\ref{theorem1}).
\end{proof}

Letting $\{\sigma_i\}$ the roots of $\mu'(1-{s})+\rho\,k_1{\prod_{j=1}^{n_\text{I}}[1+(1-{s})/k_j]}$, we have
\begin{equation}
\begin{split}
{x} =	& \mu' \sum_{i=1}^{n_\text{I}} \left(\sigma_i-1\right) \phi_i \ln\left|1-\frac{{s}}{\sigma_i}\right|\\
			& -\left(1+\rho\right) \sum_{i=1}^{n_\text{D}} \frac{1}{\kappa_i}\left(p_i^0\,{s}+\frac{{s}^2}{2}\right)\!,
\end{split}
\label{solution-gal}
\end{equation}
\noindent where
\begin{equation}
	\phi_i= \frac{\displaystyle\prod_{j=1}^{n_\text{I}} k_j}{\psi\,\displaystyle\prod_{\begin{subarray}{c}j=1\\j\ne i\end{subarray}}^{n_\text{I}}(\sigma_j-\sigma_i)} \left(1 +\sum_{j=1}^{n_\text{D}}\frac{\sigma_i+p_j^0}{\kappa_j}\right)\!.
	\label{def-phi}
\end{equation}
\noindent One should note that the $\{\sigma_i\}$ do not depend on cell death.

In Eq.~(\ref{dX_dS-death_inhib}) the numerator of the fraction is of degree 2 (of degree 1 in the absence of cell death), the denominator is of degree $n$, with $n\ge1$.
Equation~(\ref{solution-gal}) applies only if the degree of the numerator is smaller than that of the denominator (theorem~1). In the absence of cell death, Eq.~(\ref{solution-gal}) holds if there is at least one inhibition and two inhibitions or more are necessary if cell death is taken into account.

Let us consider one inhibition (parameter $k_2$) and no cell death.  If $k_2$ large in absolute value, ${s}_1 \sim k_1\,k_2/\psi$, ${{s}_2 \sim 1+\psi}$, and $\phi_i \sim (-1)^i$. Thus ${(\sigma_i-1)}\,\phi_1 {\ln |1-{s}/{{s}_1}|} \sim {s}$ and one recovers Eq.~(\ref{dS_dX-2}). It is not possible to write ${x}= \mu' {({s}_1-1)}\,\phi_1 \ln\left|1-{{s}}/{{s}_1}\right|$ because Eq.~(\ref{solution-gal}) applies only for at least one inhibition.

\section*{Applications}
Equation~(\ref{solution-gal}) is very general as it can account for an arbitrary number of both substrate and product inhibitions as well as biomass death. In this section we consider particular cases: inhibition without death, cell death without inhibition before combining the two effects.

\begin{figure}
\centering
	\includegraphics[width=7.5cm]{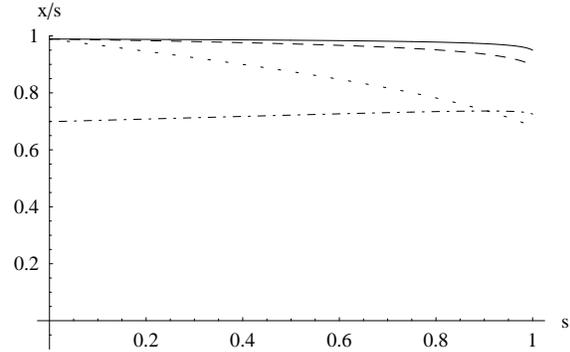}
\caption{\label{fig-inhib} Normalized biomass yield, ${x}/{s}$, as a function of ${s}$, accounting for inhibition. $k_2=-1.5$ (dashed line), $k_2=-1.05$ (dotted line), $k_2=0.05$ (dot-dashed line), and $k_2 \to \pm\infty$ (i.e.\ no inhibition, solid line).
}
\end{figure}

\begin{figure}
\centering
	\includegraphics[width=7.5cm]{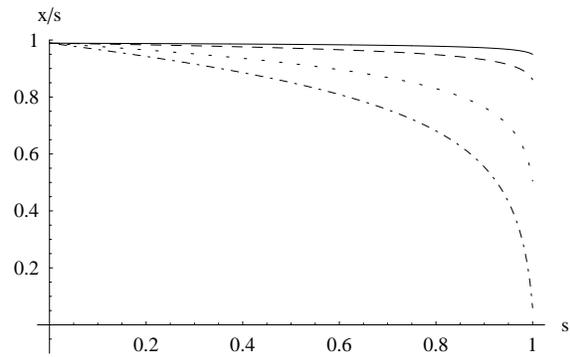}
\caption{\label{fig-death} Normalized biomass yield, ${x}/{s}$, as a function of ${s}$, accounting for cell death. $\kappa$ is  $0.5$ (dashed line), $0.1$ (dotted line), and $0.05$ (dot-dashed line). The solid line corresponds to the absence of cell death ($\kappa \to \infty$).}
\end{figure}

\begin{figure}
\centering
	\includegraphics[width=7.5cm]{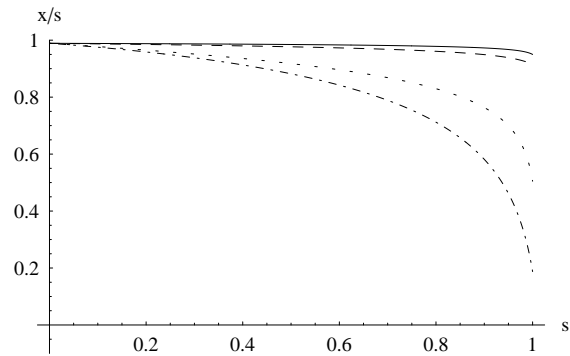}
\caption{\label{fig-death_inhib} Normalized biomass yield, ${x}/{s}$, as a function of ${s}$. The solid line corresponds to the absence of inhibition and cell death. The dashed line is for product inhibition alone ($k_2=-1.5$) and the dotted line for biomass death alone ($\kappa=0.1$). The dot-dashed line shows the combination of the two effects.}
\end{figure}

Figure~\ref{fig-inhib} shows ${x}/{s}$ in the case of one inhibition (parameter $k_2$). 
As expected substrate inhibition (dot-dashed line) affects the fermentation from the beginning whereas product inhibition matters at the end of fermentation. The total yields for $k_2=0.05$ (dot-dashed line) and $k_2=-1.05$ (dotted line) are similar although the routes are quite different. $k_2$ cannot be between $-1$ and 0 because of its definition, $k_2=-1$ corresponds to $K'=0$ if one writes $(1-P/K')^{-1}$ instead of $(1-S/K_2)^{-1}$. 
The values of the constants used in Figs.~\ref{fig-inhib} and subsequent are \cite{Topiwala-71}: $\mu = 0.11$~h$^{-1}$, $m = 0.01$~h$^{-1}$, $Y = 0.1$, and $K = 112$~g/L. The initial substrate concentration is $S^0=100$~g/L and $\alpha=1/2$. This gives $\psi \approx 1~\%$. 

Figure~\ref{fig-death} shows ${x}/{s}$ as a function of ${s}$ if cell death occurs but there is no inhibition. If $\kappa$ is small, ${s}$ may not reach 1 (stuck fermentation) as shown by the dot-dashed line in Fig.~\ref{fig-death}. \citet{Cramer-01} found a value for $K_\text{D}$ close to 5~kg/L~h which corresponds to ${\kappa \approx 0.1}$.

Figure~\ref{fig-death_inhib} shows ${x}/{s}$ as a function of ${s}$ when both product inhibition (parameter $k_2$) and cell death (parameter $\kappa$) occur. The initial product concentrations, $\{p_i^0\}$, are all set to zero.
The effects of product inhibition and cell death do not simply add up. Product inhibition alone (dashed line) reduces the yield only slightly. However the yield decreases significantly when product inhibition exists on top of cell death (dot-dashed line compared to dotted line). The same final value of ${x}$ can be obtained with $\kappa=0.1$ and $k_2=-1.5$ or with $\kappa\approx0.049$ and no inhibition.
The dot-dashed line in Fig.~\ref{fig-death_inhib} ($\kappa=0.1$ and $k_2=-1.5$) is indeed very similar to the dot-dashed line in Fig.~\ref{fig-death} ($\kappa=0.05$, no inhibition).

\section*{Conclusion}
By eliminating time from the kinetic equations, one can solve them in phase space, i.e.\ the biomass concentration is obtained as a function of the substrate concentration. It is then possible to obtain an analytical expression for the effect of an arbitrary number of substrate and product inhibitions combined with biomass death, without assuming constant yield. Once parameters are obtained for a certain process, substrate and product concentrations can be obtained analytically from the measured biomass concentration in order to follow the evolution of the fermentation.

\section*{Nomenclature}
\vspace{-.45cm}\noindent{\footnotesize
\begin{tabbing}
$K_i$ \:	\= inhibition coefficient \hspace{11.em}			\= (g/L)\\
$k_i$ 		\> dedimensionalized $K_i$, $k_i = K_i/S^0$			\> (-)\\
$K_\text{D}$\> cell death coefficient 							\> (g/L\,h)\\
$m$			\> maintenance coefficient 							\> (h$^{-1}$)\\
$n_\text{D}$\> number of biomass death terms					\> (-)\\
$n_\text{I}$\> number of inhibitions plus one					\> (-)\\
$P_i$		\> product concentration 							\> (g/L)\\
$p_i$		\> dedimensionalized $P_i$, $p_i=P_i/(\alpha_i S^0)$	\> (-)\\
$S$			\> substrate concentration 							\> (g/L)\\
${s}$		\> normalized substrate, ${s}=1-S/S^0$					\> (-)\\
$t$			\> time 											\> (h)\\
$X$			\> viable biomass concentration						\> (g/L)\\
${x}$		\> normalized biomass, $x\!=\!(X\!\!-\!X^0)(1\!+\!\rho)/(Y S^0)$	\> \;(-)\\
$Y$			\> biomass yield									\> (-)\\
$\alpha_i$	\> stoichiometric coefficient for product $P_i$		\> (-)\\
$\kappa$	\> dedimensionalized $K_\text{D}$, $\kappa=m Y K_\text{D}/(\alpha S^0)$\> (-)\\
$\mu$		\> \emph{maximum} specific growth rate 				\> (h$^{-1}$)\\
$\mu'$		\> change in $\mu$ due to inhibition 				\> (-)\\
$\rho$		\> ratio of time constants, $\rho = m\,Y/\mu$		\> (-)\\
$\sigma_i$	\> a root of $\mu'(1-s)+\rho\,k_1{\prod_j[1+(1-s)/k_j]}$		\> (-)\\
$\phi_i$	\> defined in Eq.~(\ref{def-phi}) 					\> (-)\\
$\psi$		\> distance to constant yield, $\psi = k_1\,\rho/(1+\rho)$ \> (-)\\
\\
{\itshape Superscripts}\\
0		\> initial value\\
\end{tabbing}
}\vspace{-.5cm}

\bibliography{fermentation}
\bibliographystyle{apsrev}

\end{document}